\documentclass{article}
\usepackage{spconf,amsmath,graphicx}

\usepackage{cite}
\usepackage{amssymb,amsfonts}
\usepackage{textcomp}
\usepackage{xcolor}
\def\BibTeX{{\rm B\kern-.05em{\sc i\kern-.025em b}\kern-.08em
    T\kern-.1667em\lower.7ex\hbox{E}\kern-.125emX}}
\usepackage{mathtools}
\usepackage{amsthm}
\usepackage{stfloats}
\usepackage{epstopdf}
\usepackage[linesnumbered,ruled,vlined]{algorithm2e}
\usepackage{subcaption}
\usepackage{caption}
\usepackage{algcompatible}
\usepackage{epstopdf}

\usepackage{lipsum}

\newcommand\blfootnote[1]{%
  \begingroup
  \renewcommand\thefootnote{}\footnote{#1}%
  \addtocounter{footnote}{-1}%
  \endgroup
}

\title{Low complexity equalization for AFDM in doubly dispersive channels}

\name{Ali Bemani$^{1}$, Nassar Ksairi$^{2}$, Marios Kountouris$^{1}$}
\address{$^{1}$ Communication Systems Department, EURECOM, Sophia Antipolis, France\\
$^{2}$ Mathematical and Algorithmic Sciences Lab, Huawei France R\&D, Paris, France \\
Email: ali.bemani@eurecom.fr, nassar.ksairi@huawei.com, marios.kountouris@eurecom.fr}

\begin{document}
%
\maketitle

\begin{abstract}
Affine Frequency Division Multiplexing (AFDM), which is based  on  discrete  affine  Fourier  transform (DAFT), has recently been proposed for reliable communication in high-mobility scenarios. Two low complexity detectors for AFDM are introduced here. Approximating the channel matrix as a band matrix via placing null symbols in the AFDM frame in the DAFT domain, a low complexity MMSE detection is proposed by means of the $\rm{LDL}$ factorization. Furthermore, exploiting the sparsity of the channel matrix, we propose a low complexity iterative decision feedback equalizer (DFE) based on weighted maximal ratio combining (MRC), which extracts and combines the received multipath components of the transmitted symbols in the DAFT domain. Simulation results show that the proposed detectors have similar performance, while weighted MRC-based DFE has lower complexity than band-matrix-approximation LMMSE when the channel impulse response has gaps.
\end{abstract}
\begin{keywords}
AFDM, affine Fourier transform, doubly dispersive channels, detector, MMSE, DFE, MRC.
\end{keywords}
\vspace{-2mm}
\section{Introduction}
\vspace{-2mm}
\blfootnote{This work has been supported by a Huawei France-funded Chair towards Future Wireless Networks.} Next-generation wireless systems (e.g., B5G/6G) are evolving to cater to a wide range of applications and services requiring reliable communication in high-mobility scenarios. This calls for new waveform design able to cope with time-varying channels. In this setting, existing waveforms, in particular orthogonal frequency  division  multiplexing (OFDM), lose subcarrier orthogonality, thus resulting in inter-carrier interference and deteriorated system performance. 

Affine frequency division multiplexing (AFDM) has recently been proposed as a promising waveform for communication in time-varying channels \cite{bemani2021afdm, bemani2021afdmhighfreq} showing significant performance gains over OFDM.  
AFDM employs multiple orthogonal information-bearing chirps generated using the discrete affine Fourier transform (DAFT). A key feature is that its chirp pulse parameters can be adapted to the channel characteristics, making a complete delay-Doppler representation of the channel in the DAFT domain. This enables AFDM to achieve full diversity in doubly dispersive channels \cite{bemani2021afdm} as opposed to existing chirp-based waveforms, for instance \cite{bomfin2021robust,bomfin2019low}. Furthermore, AFDM has similar performance in terms of bit error rate (BER) with orthogonal time frequency space (OTFS) \cite{hadani2017orthogonal}. However, AFDM outperforms OTFS in terms of pilot overhead and multiuser multiplexing overhead \cite{bemani2021afdmhighfreq,raviteja2019embedded}.

In this paper, we propose two low complexity detection algorithms for AFDM taking advantage of its inherent channel sparsity. By placing some null symbols - zero padding the AFDM frame - in the DAFT domain, the channel matrix can be approximated as a band matrix. Using the approximated band matrix and inspired by the equalizer in \cite{rugini2005simple} for OFDM systems, we first design a low complexity MMSE detector based on $\rm{LDL}$ factorization \cite{golub1996matrix}. The overall complexity of the proposed algorithm is linear in the number of subcarriers and quadratic in the bandwidth of the band matrix, which in turn depends on the the maximum delay and maximum Doppler shift. Second, we propose a low complexity iterative decision feedback equalizer (DFE) based on weighted maximal ratio combining (MRC) of the channel impaired input symbols received from different paths. The overall complexity of the second algorithm is also linear in the number of subcarriers and quadratic in the number of paths. We show that these two detectors have similar performance between them, whereas when the channel is sparse in the delay domain, weighted MRC-based DFE detector exhibits lower complexity.

\vspace{-4mm}
\section{System Model}
\begin{figure}
  \centering
  \includegraphics[scale=.46]{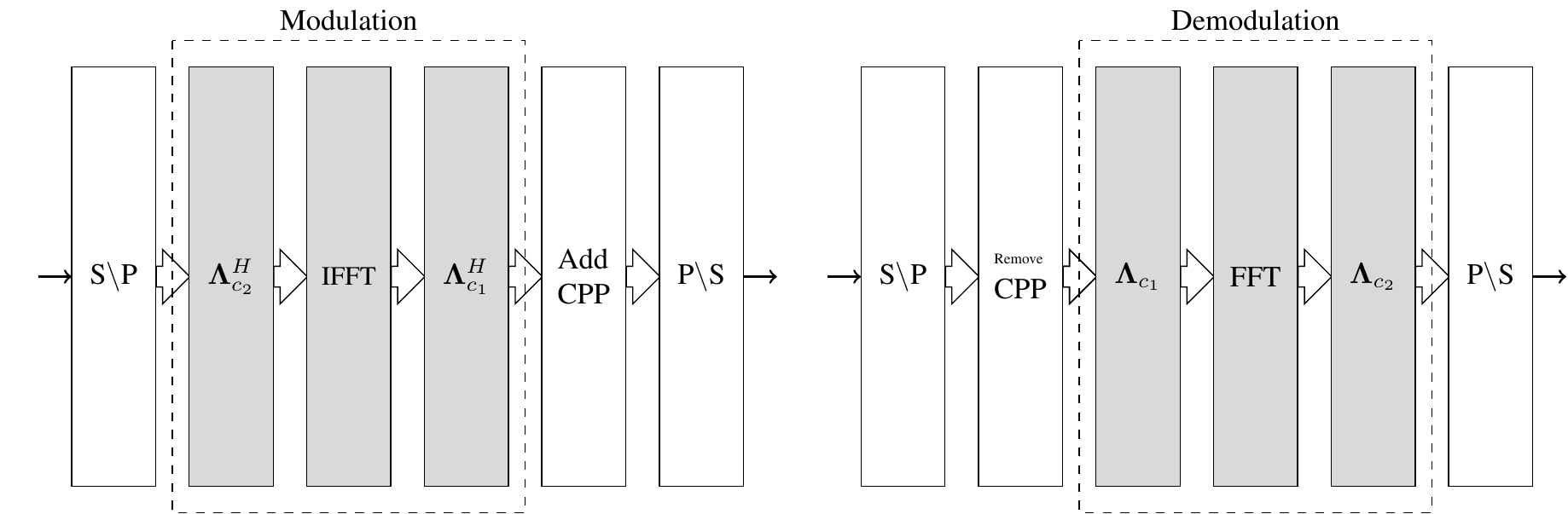}
  \caption{AFDM modulation/demodulation block diagram.}
  \vspace{-4mm}
  \label{fig:AFDM_blokcdiagrma}
\end{figure}

\vspace{-3mm}
\label{sec:afdm}
The AFDM block diagram is given in Fig.~\ref{fig:AFDM_blokcdiagrma}. Modulation is produced by using DAFT at the transmitter and receiver. DAFT is a discretized version of AFT \cite{bemani2021afdm,healy2015linear, pei2001relations, pei2000closed} and its kernel is equal to $e^{-\imath2\pi (c_2m^2+{\frac{1}{N} }mn+c_1n^2)}$ where $c_1$ and $c_2$ are the AFDM parameters tuned to provide full delay-Doppler representation of the channel in the DAFT domain. It has been shown that tuning $c_1$ using the the maximum Doppler shift normalized with respect to the subcarrier spacing, and setting $c_2$ to be an arbitrary irrational number or a rational number sufficiently smaller than ${1}/{2N}$, enables AFDM to achieve full diversity in doubly dispersive channels \cite{bemani2021afdm}. 

\vspace{-2mm}
\subsection{Modulation and Demodulation}
\vspace{-2mm}
Consider a set of quadrature amplitude modulation (QAM) symbols $x_k, k = 0, 1, 2, ..., N-1$. AFDM maps $x_k$ to $s_n$ using inverse DAFT (IDAFT) as follows:
{\small{\begin{equation}
\label{eq:mod}
    s_n = \frac{1}{\sqrt{N}}\sum_{m = 0}^{N-1}x_me^{\imath2\pi (c_2m^2+{\frac{1}{N}}mn+c_1n^2)}, \quad n= 0 , \cdots, N-1.
\end{equation}}}
To make the channel lie in a periodic domain, a  {\emph{chirp-periodic prefix}} (CPP) should be added to the modulated signal, defined as
\vspace{-2mm}
\begin{equation}
    s_{n} = s_{N+n}e^{-\imath2\pi c_1(N^2+2Nn)},\quad n = -M, \cdots, -1
\end{equation}
where $M$ is any integer greater than or equal to the value in samples of the maximum delay spread of the wireless channel. After transmission over the channel, the received samples are 
\vspace{-2mm}
\begin{equation}
    r_n = \sum_{l = 0}^{\infty}s_{n-l}g_n(l)\label{r_n} + w_n
\end{equation}
where $w_n\sim\mathcal{CN}\left(0,N_0\right)$ is an additive Gaussian noise and $g_n(l)$ is the impulse response of the time-varying channel at time $n$ and delay $l$, given by
\vspace{-4mm}
\begin{equation}
    g_n(l) = \sum _{i=1}^{P} h_{i}e^{-\imath2\pi f_in}\delta(l - l_i)
    \vspace{-2mm}
\end{equation}
 where $P\geq1$ is the number of paths, $\delta(\cdot)$ is the Dirac delta function, and $h_i, f_i$ and $l_i$ are the complex gain, Doppler shift (in digital frequencies), and the integer delay associated with the $i$-th path, respectively. We define $\nu_i \triangleq{} N f_i$, where $\nu_i\in\left[-\nu_{\max},\nu_{\max}\right]$ is the Doppler shift normalized with respect to the subcarrier spacing. We assume that the maximum delay of the channel satisfies $l_{\max}\triangleq\max (l_i)  < N$.\\
The DAFT domain output symbols are obtained by
\begin{equation}
    y_m = \frac{1}{N}\sum_{n = 0}^{N-1}r_ne^{-\imath2\pi (c_2m^2+{\frac{1}{N}}mn+c_1n^2)}.
    \label{eq:y_received}
\end{equation}
Discarding the CPP, the input-output relation can be written in matrix form as
\begin{equation}
\begin{aligned}
    {\mathbf y} =& \mathbf{A}\mathbf{r}
    = {\mathbf H}_{\mathrm{eff}} {\mathbf x}+   \mathbf{A}{\mathbf w}\label{eq_rec}
\end{aligned}
\end{equation}
where $ \mathbf{A} = {\mathbf \Lambda}_{c_2}{\mathbf F}{\mathbf \Lambda}_{c_1}$ is the DAFT matrix, $\mathbf{F}$ is the discrete Fourier transform (DFT) matrix with entries $e^{-\imath2\pi mn/N}/\sqrt{N}$, $\mathbf{\Lambda}_{c}= {\rm diag} (e^{-\imath2\pi cn^{2}}, n=0, 1, \, \ldots\,, N-1)$, ${\mathbf H}_{\mathrm{eff}}=\mathbf{A}\mathbf{H} \mathbf{A}^{\rm{H}}$, and ${\mathbf H}$ is the matrix representation of the channel. The elements of $\mathbf{y}$, $\mathbf{r}$, $\mathbf{x}$ and $\mathbf{w\sim\mathcal{CN}\left(\mathbf{0},N_0\mathbf{I}\right)}$ are similarly related to $y_k$, $r_k$, $x_k$ and $w_k$, respectively. Since $\mathbf{A}$ is a unitary matrix, $\widetilde{\mathbf{w}} = \mathbf{A}{\mathbf w}$ and ${\mathbf w}$ have the same statistics. ${\bf H}_{\rm eff}$ has sparse structure i.e., there are exactly $L=P$ non-zero entries in each row and column if $\nu_i$ is integer and if we set $c_1=\frac{2\nu_{\rm max}+1}{2N}$. Indeed, in this case ${\bf H}_{\rm eff}=\sum_{i=1}^{P}h_i{\bf H}_i$ where
{\small
\begin{equation}
    \mathbf{H}_i(p, q) = 
    \begin{cases} e^{\imath\frac{2\pi}{N}(Nc_1l_i^2 - ql_i + Nc_2(q^2 - p^2))} & q = (p+\mathrm{loc}_i)_N  \\
    0 & {\text{otherwise}}
    \end{cases}
    \label{eq:Hi_p_q_integer}
\end{equation}
}where $\mathrm{loc}_i=(\nu_i+(2\nu_{\rm max}+1)l_i)_N$\cite{bemani2021afdm}.
For fractional $\nu_i$, it can be shown that $\mathbf{H}_i$ has in each row and column a peak surrounded by approximately $2k_{\nu}$ non-zero entries decreasing rapidly as we move away from it. Hence, if we set $c_1=\frac{2(\lfloor\nu_{\rm max}\rfloor+k_{\nu})+1}{2N}$, $\mathbf{H}_{\textrm{eff}}$ can be approximated as having $L=(2k_{\nu}+1)P$ non-zero entries per row and column (i.e., each received symbol can be approximately expressed as a linear combination of only a few input symbols). We propose below two low complexity detection algorithms leveraging the channel matrix sparsity.

\vspace{-2mm}
\section{Detection Algorithms}\label{sec_detection}
\begin{figure}
  \centering
  \includegraphics[scale=.45]{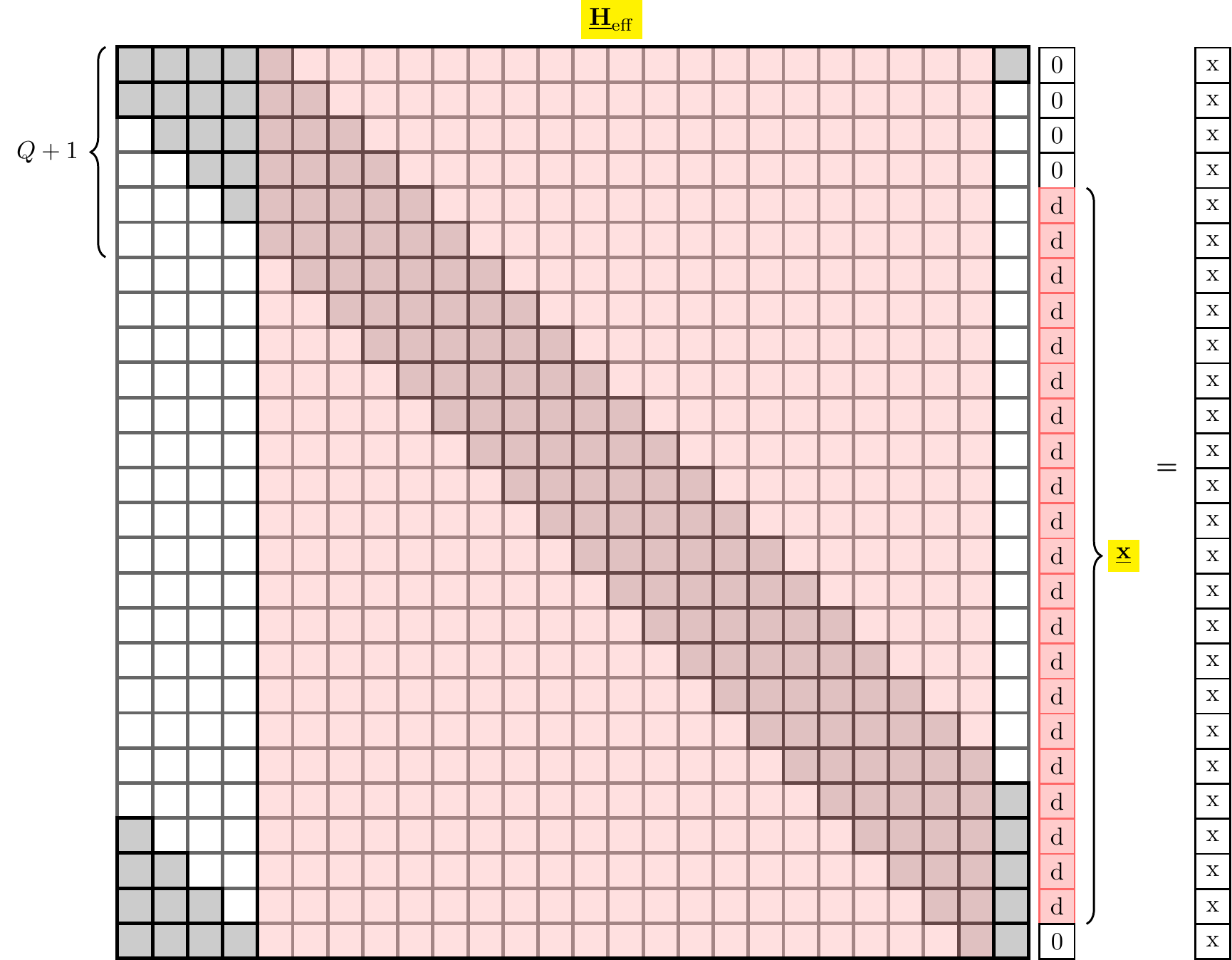}
  \caption{Truncated parts of ${\mathbf x}$ and $\mathbf{H}_{\mathrm{eff}}$}
  \vspace{-4mm}
  \label{fig:ChannelForMMSE_withoutpilot}
\end{figure}

The first step is to place some null symbols that allow to approximate the truncated part of $\textbf{H}_{\rm{eff}}$ as a band matrix. This also simplifies the input-output relation as the modular operation is no longer needed, as shown in Fig.~\ref{fig:ChannelForMMSE_withoutpilot}. Note that these symbols do not entail extra overhead as they can serve not only the proposed detection algorithms but also embedded pilot aided channel estimation. Due to the structure of ${\bf H}_{\rm eff}$ and ${\bf H}_i$, the number of the null guard symbols should be greater than $Q = (l_{\max} + 1)(2(\alpha_{\max} + k_\nu) + 1)-1$. Taking into account the zero padding, the vector of DAFT domain received samples writes as
\vspace{-2mm}
\begin{equation}
    {\mathbf y} = \underline{\mathbf H}_{\mathrm{eff}} \underline{\mathbf x}+ \widetilde{\underline{\mathbf w}}\label{eq:detectionEq}
\end{equation}
where $\underline{\mathbf x}$ and $\underline{\mathbf H}_{\mathrm{eff}}$ are the truncated parts of ${\mathbf x}$ and $\mathbf{H}_{\mathrm{eff}}$, respectively (see Fig. \ref{fig:ChannelForMMSE_withoutpilot}). They can be expressed using the matrix $\mathbf{T} = [\mathbf{I}_N]_{Q-(\alpha_{\max}+k_\nu):N-(\alpha_{\max}+k_\nu)-1, :}$ as $\underline{\mathbf x} = \mathbf{T}\mathbf{x}$ and  $\underline{\mathbf H}_{\mathrm{eff}} = {\mathbf H}_{\mathrm{eff}} \mathbf{T}^{\rm{H}}$.
Using LMMSE equalization based on \eqref{eq:detectionEq} for detection requires $\mathcal{O}(N^3)$ flops, which can be prohibitive for large $N$. We thus propose two detectors with lower complexity. The first is a low-complexity LMMSE based on a band approximation of $\underline{\mathbf H}_{\mathrm{eff}}$. The second is a weighted MRC-based DFE exploiting the sparse representation of the communication channel provided by AFDM.

\vspace{-4mm}
\subsection{Low complexity MMSE detection}
To recover the data symbols $\underline{\mathbf{x}}$, considering \eqref{eq:detectionEq}, the following MMSE equalization is used
\begin{equation}
    \hat{\underline{\mathbf{x}}} = \underline{{\mathbf H}}_{\mathrm{eff}}^{\rm{H}}(\underline{{\mathbf H}}_{\mathrm{eff}}\underline{{\mathbf H}}_{\mathrm{eff}}^{\rm{H}} + N_0\mathbf{I}_{N})^{-1}{\mathbf{y}}.
    \label{eq:Modified_MMSE_eq}
\end{equation}
Although \eqref{eq:Modified_MMSE_eq} involves matrix inversion, the matrix $\mathbf{M} = \underline{{\mathbf H}}_{\mathrm{eff}}\underline{{\mathbf H}}_{\mathrm{eff}}^{\rm{H}}+ N_0\mathbf{I}_{N}$ is a Hermitian band matrix with lower and upper bandwidth $Q$. Thus, $\mathbf{M}^{-1}$ can be computed using $\rm{LDL}$ factorization. Algorithm \ref{algo:MMSE_detect} can be performed to efficiently equalize the received signal.
\vspace{-4mm}
\begin{algorithm}
Construct the matrix $\underline{\mathbf H}_{\mathrm{eff}} ={\mathbf H}_{\mathrm{eff}} \mathbf{T}^{\rm{H}}$\\
    Construct the band matrix $\mathbf{M} = \underline{{\mathbf H}}_{\mathrm{eff}}\underline{{\mathbf H}}_{\mathrm{eff}}^{\rm{H}}+ N_0\mathbf{I}_{N}$\\
    Compute the $\rm{LDL}$ factorization of $\mathbf{M} = \mathbf{LDL}^{\rm{H}}$ where $\mathbf{L}$ is a lower triangular matrix with $Q$ sub diagonals and $\mathbf{D}$ is a diagonal matrix\\
    Solve the triangular system $\mathbf{Lf} = {\mathbf{y}}$\\
    Solve the diagonal system $\mathbf{Dg} =\mathbf{f}$\\
    Solve the triangular system $\mathbf{L}^{\rm{H}}\mathbf{d} =\mathbf{g}$\\
    Calculate $\hat{\underline{\mathbf{x}}} = \underline{\mathbf H}_{\mathrm{eff}}^{\rm{H}}\mathbf{d}$
    
\caption{Low complexity MMSE detection}
\label{algo:MMSE_detect}
\end{algorithm}
\vspace{-4mm}
The computational cost of the proposed detection algorithm is evaluated in terms of complex additions (CAs), complex multiplications (CMs) and complex divisions (CDs). The first step does not need any complex operation since $\underline{\mathbf H}_{\mathrm{eff}}$ is truncated from ${\mathbf H}_{\mathrm{eff}}$. In step 2, every element of $\underline{{\mathbf H}}_{\mathrm{eff}}\underline{{\mathbf H}}_{\mathrm{eff}}^{\rm{H}}$ requires at most $Q+1$ CMs and $Q$ CAs. Considering that $\underline{{\mathbf H}}_{\mathrm{eff}}\underline{{\mathbf H}}_{\mathrm{eff}}^{\rm{H}}$ is Hermitian and neglecting some small terms in the complexity expression, step 2 requires $\frac{1}{2}(Q^2+3Q+2)N$ CMs and $\frac{1}{2}(Q^2+Q+2)N$ CAs. Similar to \cite{rugini2005simple}, step 3, the $\rm{LDL}$ factorization, requires $\frac{1}{2}(Q^2 + 3Q)N$ CMs, $\frac{1}{2}(Q^2 + Q)N$ CAs, and $QN$ CDs. Steps 4 and 6 can be solved by band forward and backward substitutions \cite{golub1996matrix} and each of them has $QN$ CMs and $QN$ CAs. Step 5 can be solved using $N$ CDs since $\mathbf{D}$ is a diagonal matrix and the last step requires $(Q+1)N$ CMs and $QN$ CAs. Thus, the algorithm requires $(Q^2+6Q+2)N$ CMs, $(Q^2+4Q+1)N$ CAs and $(Q+1)N$ CDs, which amounts to $(2Q^2+11Q+4)N$ complex operations in total. 
\vspace{-4mm}
\subsection{Weighted MRC-based DFE detection}
\begin{figure}
  \centering
  \includegraphics[scale=.40]{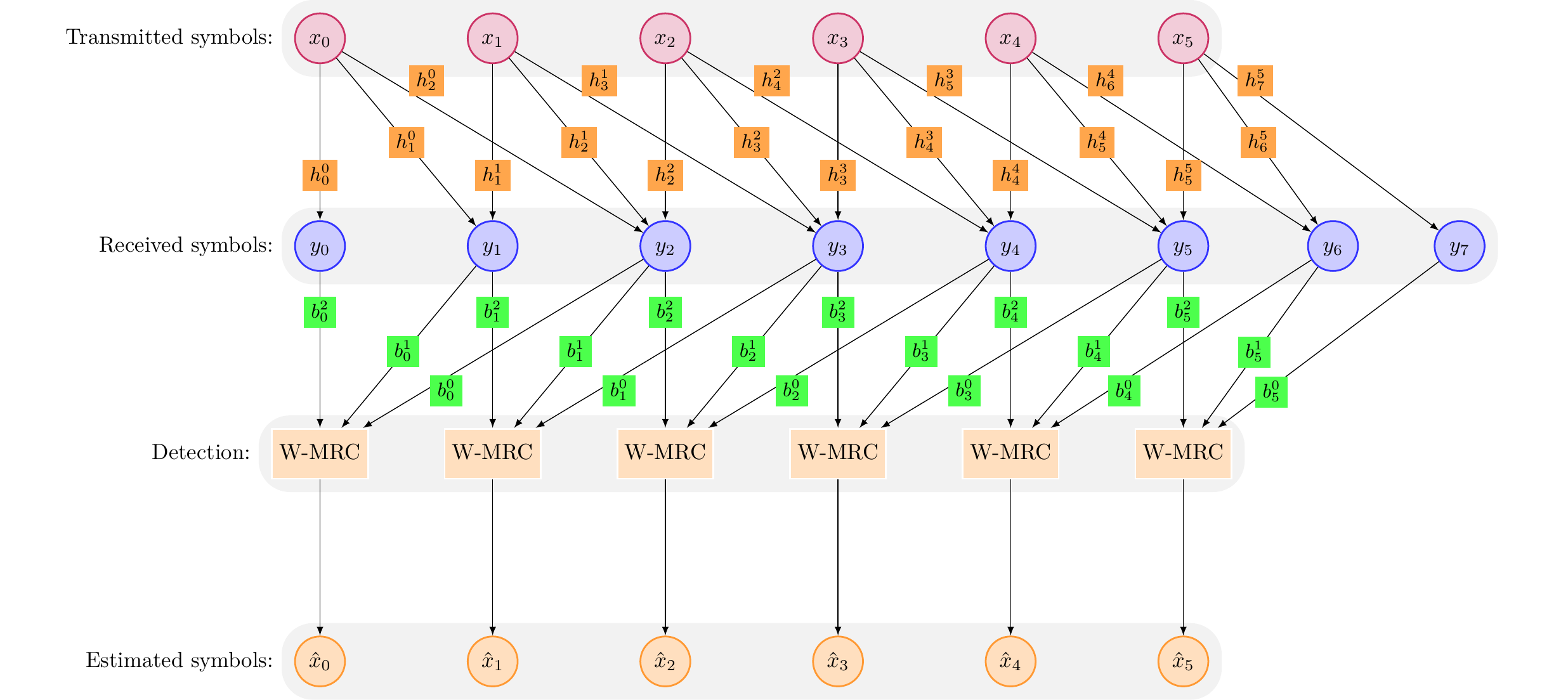}
  \caption{Weighted MRC operation for $N = 8$ with a 3-path channel with $Q=2$ where W-MRC stands for weighted MRC and $h_i^j = \underline{\mathbf H}_{\mathrm{eff}}(i, j)$.}
  \vspace{-4mm}
  \label{fig:MRC_detector}
\end{figure}

As mentioned in Section \ref{sec:afdm}, $\underline{\mathbf H}_{\mathrm{eff}}$ has $L$ non-zero entries per column. This feature enables us to propose a weighted MRC-based detector where each data symbol is detected from the weighted MRC of its $L$ channel-impaired received copies. Fig.~\ref{fig:MRC_detector} shows an example of this detector for AFDM with $N = 8$ and a 3-path channel with $Q=2$. The proposed detector is iterative, where in each iteration, the estimated inter symbol interference is canceled in the branches selected for the combining. Considering the structure of $\underline{\mathbf H}_{\mathrm{eff}}$, it can be seen that each received symbol $y_k$ is given by
\begin{equation}
    y_k = \sum_{i = 0}^{L-1}\underline{\mathbf H}_{\mathrm{eff}}(k, {p}_k^i)x_{{p}_k^i}
\end{equation}
where ${p}_k^i$ is the column index of the $i$-th path coefficient in row $k$ of matrix $\underline{\mathbf H}_{\mathrm{eff}}$. Let $b_k^i$ be the channel impaired input symbol $x_k$ in the received samples $y_{{q}_k^i}$ after canceling the interference from other input symbols, where ${q}_k^i$ is the row index of the $i$-th path coefficient in column $k$ of matrix $\underline{\mathbf H}_{\mathrm{eff}}$. In each iteration, assuming estimates of the input symbols $x_k$ are available either from the current iteration (for ${p}_{{q}_k^i}^j < k, \quad j = 0, ..., L-1$) or previous iteration (for $ {p}_{{q}_k^i}^j> k, \quad j = 0, ..., L-1$), $b_k^i$ can be written as 
\vspace{-2mm}
\begin{equation}
\begin{aligned}
    b_k^i = y_{{{q}_k^i}} &- \sum_{ {p}_{{q}_k^i}^j< k} \underline{\mathbf H}_{\mathrm{eff}}^{\rm{H}}({{{q}_k^i}}, { {p}_{{q}_k^i}^j})\hat{x}_{{p}_{{q}_k^i}^j}^{(n)}\\
    &- \sum_{ {p}_{{q}_k^i}^j> k} \underline{\mathbf H}_{\mathrm{eff}}^{\rm{H}}({{q}_k^i}, { {p}_{{q}_k^i}^j})\hat{x}_{ {p}_{{q}_k^i}^j}^{(n-1)}\label{eq:b_k_i}
\end{aligned}
\vspace{-2mm}
\end{equation}
where superscript $(n)$ denotes the $n$-th iteration.
 Considering $b_k^i$ for all paths $i = 0, 1, ..., L-1$, weighted MRC (as opposed to pure MRC \cite{thaj2020low})  can be performed. The output of the weighted MRC for estimating $x_k$ is given by
\begin{equation}
    c_k = \frac{g_k}{d_k + \gamma^{-1}}\label{eq:c_kMRC}
\end{equation}
where 
\vspace{-2mm}
\begin{equation}
    g_k \triangleq \sum_{i = 0}^{L-1}\underline{\mathbf H}_{\mathrm{eff}}^{\rm{H}}({q}_k^i, k)b_k^i,
\end{equation}
\begin{equation}
    d \triangleq \sum_{i = 0}^{L-1}|\underline{\mathbf H}_{\mathrm{eff}}^{\rm{H}}({q}_k^i, k)|^2
\end{equation}
and $\gamma$ is the signal-to-noise ratio (SNR). Let $\mathcal{D(.)}$ denote the decision on the symbol estimate $c_k$, i.e, $\hat{x}_k^n = \mathcal{D}(c_k)$. In this paper, we consider $\hat{x}_k^n = c_k$. The estimated symbols are then used for the next iteration. The algorithm continues until the maximum number of iterations is reached or the updated input symbol vector is close enough to the previous one as summarized in Algorithm \ref{algo:MRC_detection}.

\vspace{-3mm}
\begin{algorithm}
  \SetAlgoLined
  \KwData{$\underline{\mathbf H}_{\mathrm{eff}}$, $d$, $\mathbf{y}$, $\hat{\mathbf{x}}^0 = \mathbf{0}$}
   {\For{n = 1 : $n_{\rm{iter}}$}{
     \For{k = 0 : N-Q-1}{    
        \For{i = 0 : L-1}{
         \small{$\begin{aligned}
    b_k^i = y_{{{q}_k^i}} &- \sum_{ {p}_{{q}_k^i}^j< k} \underline{\mathbf H}_{\mathrm{eff}}^{\rm{H}}({{{q}_k^i}}, { {p}_{{q}_k^i}^j})\hat{x}_{{p}_{{q}_k^i}^j}^{(n)}\\
    &- \sum_{ {p}_{{q}_k^i}^j> k} \underline{\mathbf H}_{\mathrm{eff}}^{\rm{H}}({{q}_k^i}, { {p}_{{q}_k^i}^j})\hat{x}_{ {p}_{{q}_k^i}^j}^{(n-1)}
\end{aligned}$
         }
         }
         $g_k = \sum_{i = 0}^{L-1}\underline{\mathbf H}_{\mathrm{eff}}^{\rm{H}}({q}_k^i, k)b_k^i$

         $c_k = \frac{g_k}{d_k + \gamma^{-1}}$
         
         $\hat{x}_k^{(n)} = c_k$ or $\hat{x}_k^{(n)} = \mathcal{D}(c_k)$
        }
        \lIf{$||\hat{\mathbf{x}}^{(n)} - \hat{\mathbf{x}}^{(n-1)}|| < \epsilon$}{EXIT}
  }}
  \caption{Weighted MRC-based DFE detection}
  \label{algo:MRC_detection}
\end{algorithm}
\vspace{-4mm}
Computing the complexity of Algorithm \ref{algo:MRC_detection} is straightforward as it has only scalar operation. From step 3 to step 11, it requires $L^2$ CMs, $L^2$ CAs and 1 CD. Therefore, its total complexity is $n_{\rm{iter}}(2L^2+1)(N-Q)$. In simulations, we observed that the algorithm typically converges within 15 iterations. In the longer version of the article, convergence of $\hat{\mathbf{x}}^{(n)}$ to the LMMSE estimate $\hat{\underline{\mathbf{x}}}$ defined in \eqref{eq:Modified_MMSE_eq} is proved.

The complexity of the two proposed algorithms is remarkably smaller than maximum likelihood (ML) and linear MMSE detectors, which have exponential $\mathcal{O}(|\mathbb{A}|^{N})$ and cubic $\mathcal{O}(N^3)$ complexity, respectively, with $\mathbb{A}$ representing the QAM alphabet. Moreover, Algorithm \ref{algo:MRC_detection} has lower complexity than Algorithm \ref{algo:MMSE_detect} when the channel impulse response has gaps. This is due to the fact that its complexity only depends on the number of non-zero elements in each column of $\underline{\mathbf H}_{\mathrm{eff}}$, i.e $L$, instead of $Q\geq L$.

\vspace{-3mm}
\section{Simulation Results}
\begin{figure}
\centering
\begin{subfigure}{0.235\textwidth}
\centering
\includegraphics[width= \textwidth]{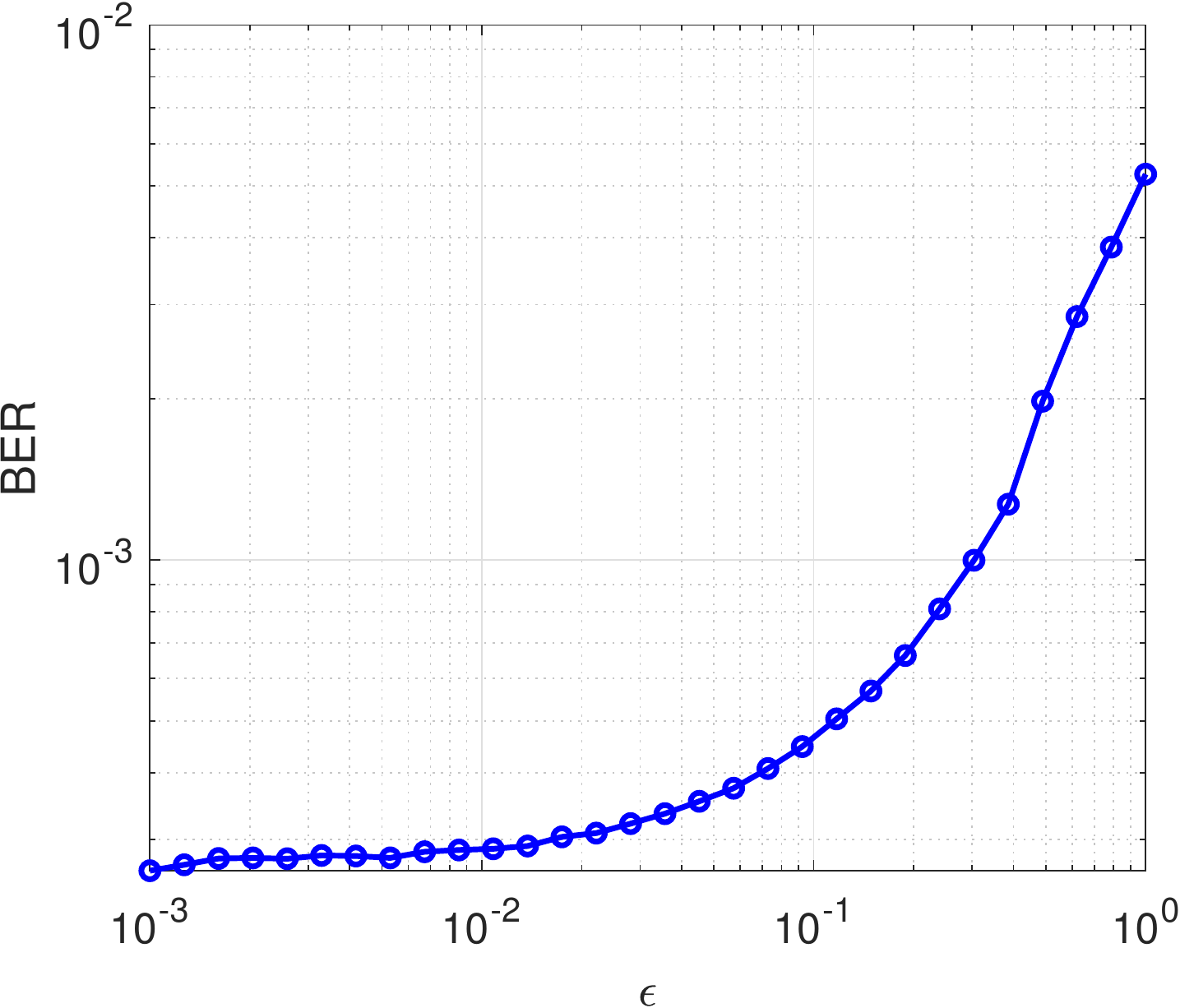}
\caption{BER variation vs. $\epsilon$.}
\label{fig:BER_varEps}
\end{subfigure}
\begin{subfigure}{0.235\textwidth}
\centering
\includegraphics[width= \textwidth]{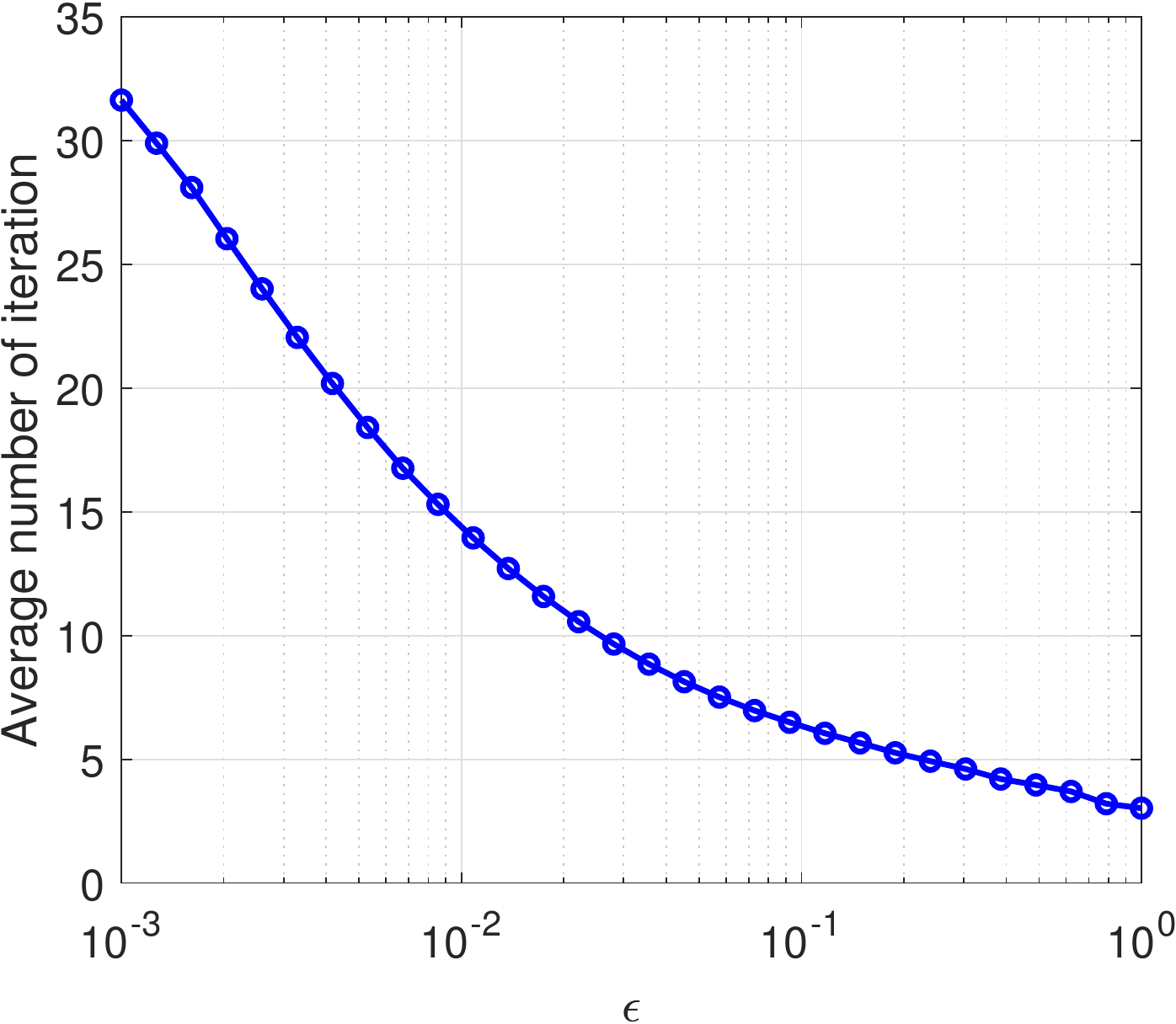}
\caption{\small{Iterations vs. $\epsilon$}.}
\label{fig:NoIterEps}
\end{subfigure}
\caption{BER variation and the average number of iterations versus $\epsilon$ for the weighted MRC-based DFE detector for 4-QAM, $N = 128$ and SNR = 20 dB.} 
\vspace{-4mm}
\label{}
\end{figure}

\begin{figure}
  \centering
  \includegraphics[scale=.55]{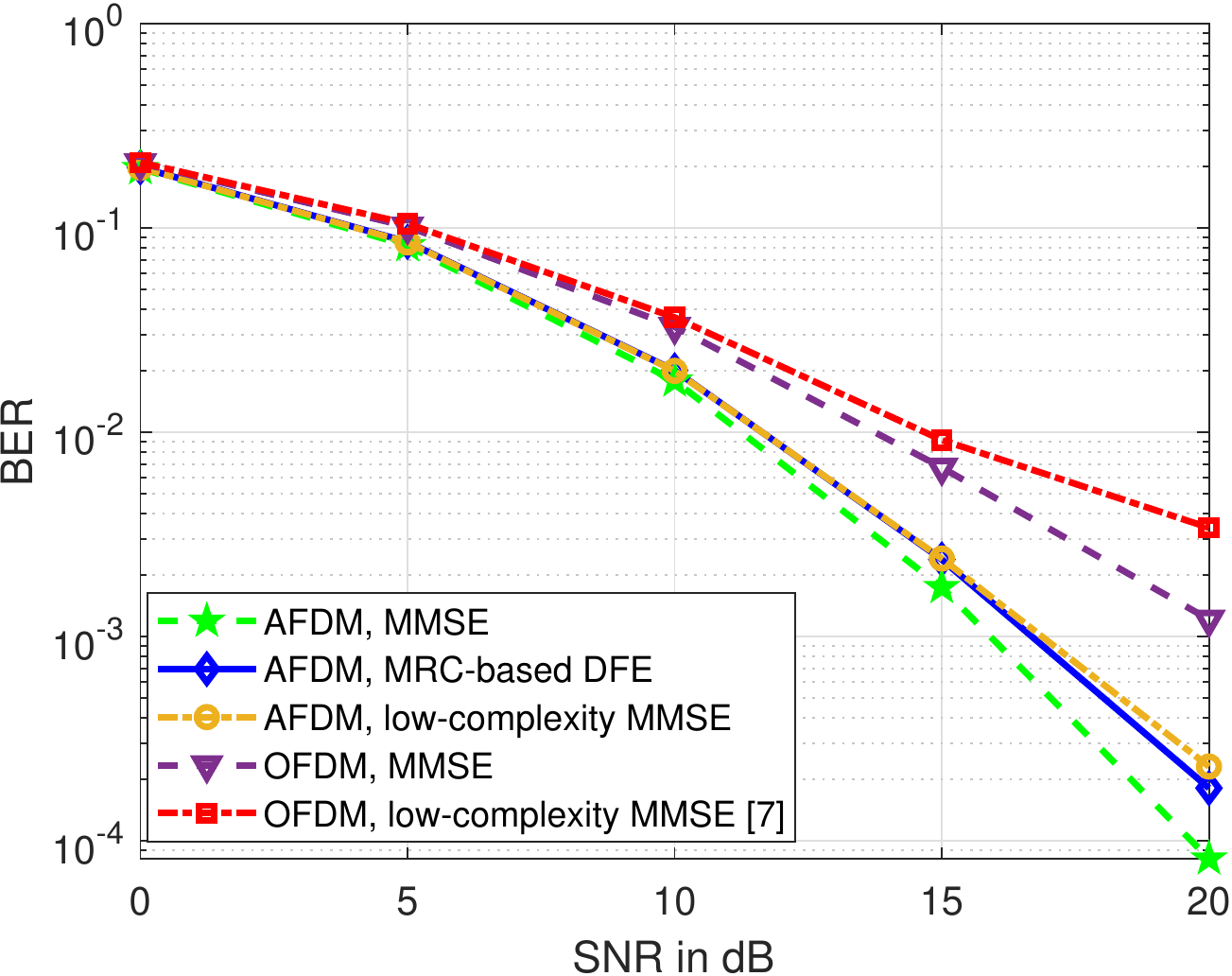}
  \caption{BER performance comparison between AFDM and OFDM systems using different detectors}
  \vspace{-5mm}
  \label{fig:MRCvsMMSE}
\end{figure}

\vspace{-2mm}
In this section, we simulate the uncoded BER performance of AFDM over doubly dispersive channels. The following parameters are used: carrier frequency $f_c= 4$ GHz, number of subcarriers $N = 128$, and AFDM frame length 330 $\mu$s. Path delays are fixed, and considering Jakes Doppler spectrum for each channel realization, the Doppler shift of the $i$-th path is generated using $\nu_i = \nu_{\max} cos(\theta_i)$, where $\theta_i$ is uniformly distributed over $[ -\pi, \pi]$ with 4-QAM signaling. The maximum Doppler shift is $\nu_{\max} = 1$, which corresponds to a maximum speed of 810 km/h.
Fig.~\ref{fig:BER_varEps} shows the BER performance of AFDM using the proposed weighted MRC-based DFE detector for different values of $\epsilon$ at SNR = 20 dB. We can see that below $\epsilon = 0.01$, the performance remains almost constant, and the algorithm converges within 14 iterations, as shown in Fig.~\ref{fig:NoIterEps}.
In Fig. \ref{fig:MRCvsMMSE}, we plot the BER performance of AFDM and OFDM for LMMSE, low-complexity MMSE \cite{rugini2005simple} and weighted MRC-based DFE detectors. First, we observe that AFDM outperforms OFDM, thanks to achieving full diversity and as every information symbol is received through multiple independent non-overlapping paths. Second, we observe that both proposed detection algorithms have similar performance between them, while conventional MMSE detection has slightly better performance at the cost of higher complexity.

\vspace{-5mm}
\section{Conclusion}
\vspace{-1.5mm}
We proposed two low complexity detection algorithms for zero-padded AFDM. First, a low complexity MMSE detector which makes use of band $\rm{LDL}$ factorization was derived. Second, an iterative weighted MRC-based DFE detector, which exploits the channel sparsity, was proposed. Our results showed that both detectors have comparable performance as exact LMMSE while their complexity order is linear, instead of cubic, in the number of subcarriers.


\bibliographystyle{IEEEtran}
\bibliography{strings,refs}

\end{document}